\documentstyle[12pt]{article}

\def\hybrid{\topmargin -20pt    \oddsidemargin 0pt
        \headheight 0pt \headsep 0pt
        \textwidth 6.25in       
        \textheight 9.5in       
        \marginparwidth .875in
        \parskip 5pt plus 1pt   \jot = 1.5ex}
\def\cQ{{\cal Q}}
\def\cG{{\cal G}}
\def\cL{{\cal L}}
\def\cH{{\cal H}}
\def\ket#1{|{#1}\rangle}
\def\noi{\noindent}
\def\half{{1\over2}}
\def\baselinestretch{1.2}

\catcode`\@=11

\def\marginnote#1{}
\def\draftlabel#1{{\@bsphack\if@filesw {\let\thepage\relax
   \xdef\@gtempa{\write\@auxout{\string
      \newlabel{#1}{{\@currentlabel}{\thepage}}}}}\@gtempa
   \if@nobreak \ifvmode\nobreak\fi\fi\fi\@esphack}
        \gdef\@eqnlabel{#1}}
\def\@eqnlabel{}
\def\@vacuum{}
\def\draftmarginnote#1{\marginpar{\raggedright\scriptsize\tt#1}}

\def\draft{\oddsidemargin -.2truein
        \def\@oddfoot{\sl preliminary draft \hfil
        \rm\thepage\hfil\sl\today\quad\militarytime}
        \let\@evenfoot\@oddfoot \overfullrule 3pt
        \let\label=\draftlabel
        \let\marginnote=\draftmarginnote
   \def\@eqnnum{(\theequation)\rlap{\kern\marginparsep\tt\@eqnlabel}%
\global\let\@eqnlabel\@vacuum}  }

\def\preprint{\twocolumn\sloppy\flushbottom\parindent 2em
        \leftmargini 2em\leftmarginv .5em\leftmarginvi .5em
        \oddsidemargin -.5in    \evensidemargin -.5in
        \columnsep .4in \footheight 0pt
        \textwidth 10.in        \topmargin  -.4in
        \headheight 12pt \topskip .4in
88      \textheight 6.9in \footskip 0pt
        \def\@oddhead{\thepage\hfil\addtocounter{page}{1}\thepage}
        \let\@evenhead\@oddhead \def\@oddfoot{} \def\@evenfoot{} }

\def\numberbysection{\@addtoreset{equation}{section}
        \def\theequation{\thesection.\arabic{equation}}}

\def\underline#1{\relax\ifmmode\@@underline#1\else
        $\@@underline{\hbox{#1}}$\relax\fi}

\def\titlepage{\@restonecolfalse\if@twocolumn\@restonecoltrue
\onecolumn
     \else \newpage \fi \thispagestyle{empty}\c@page\z@
        \def\thefootnote{\fnsymbol{footnote}} }

\def\endtitlepage{\if@restonecol\twocolumn \else \newpage \fi
        \def\thefootnote{\arabic{footnote}}
        \setcounter{footnote}{0}}  

\catcode`@=12
\relax

\def\figcap{\section*{Figure Captions\markboth
        {FIGURECAPTIONS}{FIGURECAPTIONS}}\list
        {Figure \arabic{enumi}:\hfill}{\settowidth\labelwidth{Figure
999:}
        \leftmargin\labelwidth
        \advance\leftmargin\labelsep\usecounter{enumi}}}
 \relax
\def\tablecap{\section*{Table Captions\markboth
        {TABLECAPTIONS}{TABLECAPTIONS}}\list
        {Table \arabic{enumi}:\hfill}{\settowidth\labelwidth{Table
999:}
        \leftmargin\labelwidth
        \advance\leftmargin\labelsep\usecounter{enumi}}}
 \relax
\def\reflist{\section*{References\markboth
        {REFLIST}{REFLIST}}\list
        {[\arabic{enumi}]\hfill}{\settowidth\labelwidth{[999]}
        \leftmargin\labelwidth
        \advance\leftmargin\labelsep\usecounter{enumi}}}
 \relax
\makeatletter
\newcounter{pubctr}
\def\publist{\@ifnextchar[{\@publist}{\@@publist}}
\def\@publist[#1]{\list
        {[\arabic{pubctr}]\hfill}{\settowidth\labelwidth{[999]}
        \leftmargin\labelwidth
        \advance\leftmargin\labelsep
        \@nmbrlisttrue\def\@listctr{pubctr}
        \setcounter{pubctr}{#1}\addtocounter{pubctr}{-1}}}
\def\@@publist{\list
        {[\arabic{pubctr}]\hfill}{\settowidth\labelwidth{[999]}
        \leftmargin\labelwidth
        \advance\leftmargin\labelsep
        \@nmbrlisttrue\def\@listctr{pubctr}}}
 \relax
\makeatother
\newskip\humongous \humongous=0pt plus 1000pt minus 1000pt

\newif\ifdtup

\font\Scbig=cmss10 scaled\magstep1
\font\Scscr=cmss8 scaled\magstep1
\font\Scscrscr=cmss8
\newfam\Scfam
\textfont\Scfam=\Scbig
\scriptfont\Scfam=\Scscr
\scriptscriptfont\Scfam=\Scscrscr
\def\Sc{\fam\Scfam}

\relax
\hybrid
\def\lvm{\leavevmode\hbox to\parindent{\hfill}}

\def\thefootnote{\fnsymbol{footnote}}
\def\BE{\begin{equation}}
\def\EE{\end{equation}}
\def\BA{\begin{eqnarray}}
\def\EA{\end{eqnarray}}

\def\G{\Gamma}

\def\a{\alpha}
\def\th{\theta}

\def\P{\Phi}

\def\tt{\bar\tau}

\def\lvm{\leavevmode\hbox to\parindent{\hfill}}
\def\bar{\overline}
\def\req#1{(\ref{#1})}

\def\L{\left}
\def\R{\right}

\def\BE{\begin{equation}}
\def\EE{\end{equation} \vskip 0.30\baselineskip}
\def\BA{\begin{array}}
\def\EA{\end{array}}

\def\noi{\noindent}

\def\frac#1#2{{\textstyle{{#1}\over{#2}}}}
\def\half{{1\over2}}

\def\Kr#1{\delta_{{#1},0}}
\def\ket#1{|{#1}\rangle}

\def\ccases#1#2{\L\{\!\new\BA{l}{#1}\\ {#2}\EA\R.}

\def\cA{{\cal A}}
\def\cG{{\cal G}}
\def\cH{{\cal H}}

\def\cL{{\cal L}}
\def\cO{{\cal O}}
\def\cQ{{\cal Q}}

\def\cU{{\cal U}}

\def\open#1{\mbox{{\bf{#1}}}}

\def\oZ{{\open Z}}

\def\ctop{{\Sc c}}
\def\htop{{\Sc h}}

\def\a{\alpha}
\def\b{\beta}
\def\g{\gamma}
\def\Ups{\Upsilon}
\def\kua{\ket\Upsilon^{(1)}}
\def\kub{\ket\Upsilon^{(2)}}

\newif\ifold \oldtrue \def\new{\oldfalse}
\let\ssection=\section
\def\section{\setcounter{equation}{0}\ssection}

\begin{document}
\renewcommand{\theequation}{\thesection.\arabic{equation}}
\newcommand{\beq}{\begin{equation}}
\newcommand{\eeq}[1]{\label{#1}\end{equation}}
\newcommand{\ber}{\begin{eqnarray}}
\newcommand{\eer}[1]{\label{#1}\end{eqnarray}}
\begin{titlepage}
\begin{center}

\hfill IMAFF-95/35\\
\hfill hep-th/9504056\\
\vskip .5in

{\large \bf  Spectral Flows and Twisted Topological Theories}
\vskip .8in

{\bf Beatriz Gato-Rivera and Jose Ignacio Rosado}\\
\vskip
 .3in

{\em Instituto de Matem\'aticas y F\'\i sica Fundamental, CSIC,\\ Serrano 123,
Madrid 28006, Spain} \footnote{e-mail addresses:
bgato, jirs @cc.csic.es}\\

\vskip .5in

\end{center}

\vskip .6in

\begin{center} {\bf ABSTRACT } \end{center}
\begin{quotation}
We analyze the action of the spectral flows on N=2 twisted topological
theories. We show that they provide a useful mapping between the two
twisted topological theories associated to a given N=2 superconformal
theory. This mapping can also be viewed as a topological algebra automorphism.
 In particular null vectors are mapped into null vectors, considerably
simplifying their computation. We give the level 2 results. Finally we
discuss the spectral flow mapping in the case of the DDK and KM realizations
 of the topological algebra.

\end{quotation}
\vskip 1.5cm

Revised version, August 1995\\
\end{titlepage}
\vfill
\eject
\def\baselinestretch{1.2}
\baselineskip 16 pt
\section{Introduction}\lvm

We investigate the spectral flow transformations associated to N=2
superconformal theories, when they are applied to the topological
theories obtained by twisting the superconformal ones.
Only spectral flows with $\th=\pm 1$ provide a
mapping between the two twisted topological theories associated to a
given superconformal theory; this is due to the fact that only for
those values the Hilbert spaces of the two twisted theories are
mapped into each other. Although for no value of $\th$ (except the
trivial) the theories map back to themselves, the mapping interpolating
between the two twisted theories gives rise to a topological
algebra automorphism acting inside a given theory.
 In both cases we find that null vectors are
mapped into null vectors, either from one twisted theory to the other or
inside one of the theories.
 As a result the computation of null vectors
reduces considerably, not only because there are less vectors in number
 to compute but also because hard to obtain null vectors are
directly related to much easier ones.
In section 3 we write down all the
 relevant expressions for level 2; that is, the different kinds of
descendants and null vectors, and the spectral flow relations
between them.
 In section 4 we study the spectral flow
mappings for the DDK and KM realizations of the topological algebra,
which are the two twistings of the same N=2 superconformal
theory. The DDK realization is a bosonic string construction
(with the Liouville field and $c=-26$ reparametrization ghosts),
while the KM realization is related to the KP hierarchy through
the Kontsevich-Miwa transformation. We obtain the spectral flow
mapping relating the different field components between these two
theories, finding that it does not mix fields of different nature.
As a result the null vectors of the reduced conformal
field theories (matter + scalar systems, without the ghosts) are
mapped into each other as well. Section 5 is devoted to conclusions.

\section{Spectral Flow Mappings}\lvm

The topological algebra obtained by twisting the N=2 superconformal
algebra  \cite{[EY]}, \cite{[W-top]}, \cite{DVV} reads

\BE\new\BA{lclclcl}
\L[\cL_m,\cL_n\R]&=&(m-n)\cL_{m+n}\,,&\qquad&[\cH_m,\cH_n]&=
&{\ctop\over3}m\Kr{m+n}\,,\\
\L[\cL_m,\cG_n\R]&=&(m-n)\cG_{m+n}\,,&\qquad&[\cH_m,\cG_n]&=&\cG_{m+n}\,,
\\
\L[\cL_m,\cQ_n\R]&=&-n\cQ_{m+n}\,,&\qquad&[\cH_m,\cQ_n]&=&-\cQ_{m+n}\,,\\
\L[\cL_m,\cH_n\R]&=&\multicolumn{5}{l}{-n\cH_{m+n}+{\ctop\over6}(m^2+m)
\Kr{m+n}\,,}\\
\L\{\cG_m,\cQ_n\R\}&=&\multicolumn{5}{l}{2\cL_{m+n}-2n\cH_{m+n}+
{\ctop\over3}(m^2+m)\Kr{m+n}\,,}\EA\qquad m,~n\in\oZ\,.\label{topalgebra}
\EE

\noi
where $\cL_m$ and $\cH_m$ are the bosonic generators corresponding
to the energy momentum tensor (Virasoro generators)
 and the topological $U(1)$ current respectively, while
$\cQ_m$ and $\cG_m$ are the fermionic generators corresponding
to the BRST current and the spin-2 fermionic current
 respectively. The "topological central
charge" $\ctop$ is the true central charge of the N=2
superconformal algebra \cite{LVW} .

This algebra is satisfied by the two sets of topological generators

\BE\new\BA{rclcrcl}
\cL^{(1)}_m&=&\multicolumn{5}{l}{L_m+\half(m+1)H_m\,,}\\
\cH^{(1)}_m&=&H_m\,,&{}&{}&{}&{}\\
\cG^{(1)}_m&=&G_{m+\half}^+\,,&\qquad &\cQ_m^{(1)}&=&G^-_{m-\half}
\,,\label{twa}\EA\EE

\noi
and

\BE\new\BA{rclcrcl}
\cL^{(2)}_m&=&\multicolumn{5}{l}{L_m-\half(m+1)H_m\,,}\\
\cH^{(2)}_m&=&-H_m\,,&{}&{}&{}&{}\\
\cG^{(2)}_m&=&G_{m+\half}^-\,,&\qquad &
\cQ_m^{(2)}&=&G^+_{m-\half}\,,\label{twb}\EA\EE

\noi
corresponding to the two possible twistings of the superconformal generators
$L_m, H_m, G^{+}_m$ and $G^{-}_m$. We see that $G^{+}$ and $G^{-}$ play
mirrored roles with respect to the definitions of $\cG$ and $\cQ$. In
particular $(G^{+}_{1/2}, G^{-}_{-1/2})$ results in
$(\cG^{(1)}_0, \cQ^{(1)}_0)$, while
 $(G^{-}_{1/2}, G^{+}_{-1/2})$ gives
$(\cG^{(2)}_0, \cQ^{(2)}_0)$, so that the topological primary chiral fields
$\P^{(1)}$ and $\P^{(2)}$ come from the antichiral and the chiral rings
respectively.

The spectral flows act on the superconformal generators in the following way
\cite{SS}, \cite{LVW}

\BE\new\BA{rclcrcl}
\cU_\th \, L_m \, \cU_\th^{-1}&=& L_m
 +\th H_m + {\ctop\over 6} \th^2 \delta_{m,0}\,,\\
\cU_\th \, H_m \, \cU_\th^{-1}&=&H_m + {\ctop\over3} \th \delta_{m,0}\,,\\
\cU_\th \, G^+_r \, \cU_\th^{-1}&=&G_{r+\th}^+\,,\\
\cU_\th \, G^-_r \, \cU_\th^{-1}&=&G_{r-\th}^-\,.\
\label{spfl} \EA\EE

\noi
This translates at the level of the topological generators
\req{twa} and \req{twb} into the transformations

\BE\new\BA{rclcrcl}
\cU_\th \, \cL^{(2)}_m \, \cU_\th^{-1}&=& \cL_m^{(1)}
 +(\th - (m+1))\cH_m^{(1)} + {\ctop\over6}\th(\th-(m+1)) \delta_{m,0}\,,\\
\cU_\th \, \cH^{(2)}_m \, \cU_\th^{-1}&=&-\cH_m^{(1)} -
 {\ctop\over3} \th \delta_{m,0}\,,\\
\cU_\th \, \cQ^{(2)}_m \, \cU_\th^{-1}&=&\cG_{m-1+\th}^{(1)}\,,\\
\cU_\th \, \cG^{(2)}_m \, \cU_\th^{-1}&=&\cQ_{m+1-\th}^{(1)}\,,\\
\label{spflg} \EA\EE

\noi
and exactly the same expressions exchanging
$(1) \leftrightarrow (2)$ and $\th \leftrightarrow -\th$.
 We could have expressed the right
hand side of these transformations in terms of the generators
(2) instead (with different expressions, of course).
 However, this is not of much use, as we will now
discuss. First of all, it is clear that $\th$ must be an integer
for the transformation to make any sense. Moreover, $\th$ must
be equal to 1. The reason is that only $\cU_1$ maps the
chiral ring, on which the generators (2) act, into the
antichiral ring, on which the generators (1) act. In addition, for
no value of $\th$ (except the trivial $\th=0$) the chiral
ring gets mapped into the chiral ring, making it
completely useless to express the right hand side of
\req{spflg} in terms of the generators (2) (because only
the fields $\P^{(2)}$  are chiral primaries with respect to
these). Therefore, the only sensible spectral flow mapping for
the topological generators (2) is given by

\BE\new\BA{rclcrcl}
\cU_1 \, \cL^{(2)}_m \, \cU_1^{-1}&=& \cL_m^{(1)} - m\cH_m^{(1)}\,,\\
\cU_1 \, \cH^{(2)}_m \, \cU_1^{-1}&=&-\cH_m^{(1)}
 -{\ctop\over3} \delta_{m,0}\,,\\
\cU_1 \, \cQ^{(2)}_m \, \cU_1^{-1}&=&\cG_m^{(1)}\,,\\
\cU_1 \, \cG^{(2)}_m \, \cU_1^{-1}&=&\cQ_m^{(1)}\,.\
\label{spfla}\EA\EE

For $\th=-1$ the antichiral ring maps into the chiral ring, and
one gets the same transformation \req{spfla} exchanging
$(1) \leftrightarrow (2)$ and $\cU_1 \leftrightarrow \cU_{-1}$.

Now let us consider a descendant $\ket\Ups^{(2)}$ of the twisted
theory (2) and its image under $\cU_1$,
 that is a descendant of the
twisted theory (1), say $\ket\Ups^{(1)} = \cU_1 \ket\Ups^{(2)}$.
  From \req{spfla}  we get

\BE\new\BA{rclcrcl}
\cL_0^{(1)} \, \kua &=& \cU_1 \,
\cL_0^{(2)} \, \kub=\Delta^{(2)} \, \kua\,,\\
-(\cH_0^{(1)}+{\ctop\over3}) \, \kua &=& \cU_1 \, \cH_0^{(2)} \, \kub=
 \htop^{(2)} \, \kua\,,\\
\cQ_0^{(1)} \, \kua &=& \cU_1 \,  \cG_0^{(2)} \, \kub\,,\\
\cG_0^{(1)} \, \kua &=& \cU_1  \, \cQ_0^{(2)} \, \kub\,.\EA\EE

\noi
That is, the spectral flow $\cU_1$ leaves the conformal weight
(the level) of the descendant $\ket\Ups^{(2)}$ unchanged,
$\Delta^{(1)} = \Delta^{(2)}$, while modifying the U(1)
charge, $\htop^{(1)} = -\htop^{(2)} - \ctop/3$. Moreover, $\cQ_0^{(2)}$-
invariant $(\cG_0^{(2)}$-invariant) descendants $\ket\Ups^{(2)}$
transform into $\cG_0^{(1)}$-invariant $(\cQ_0^{(1)}$-invariant)
descendants $\ket\Ups^{(1)}$.

In addition, if $\kub$ is a null vector, the highest weight conditions
$\cL_{m > 0}^{(2)} \kub = \cH_{m > 0}^{(2)} \kub =
\cG_{m > 0}^{(2)} \kub = \cQ_{m > 0}^{(2)} \kub = 0$
result in highest weight conditions for $\kua$ too. Conversely,
if $\kua$ is a null vector, $\cU_{-1} \kua = \kub$ is a null vector
as well. Notice that the spectral flow (2.4) acting on the untwisted
N=2 superconformal algebra does not map null vectors built on the
chiral ring into null vectors built on the antichiral ring.
It even changes the level of the descendants, depending on their
$U(1)$ charges. The so called mirror map between the two twisted
topological theories \cite{RS} also fails to map null states into
null states.

At this point we have to make an observation. In dealing with topological
descendants, constructing null vectors, etc, as long as we stay at
the topological algebra level, without going into particular realizations,
we can regard the transformation \req{spfla} (without the labels (1)
 and (2)) as an internal mapping of the topological algebra
\req{topalgebra}. As a matter of fact, that transformation without the
labels is  {\it an automorphism of the topological algebra}, as the reader
can easily verify. This is a reflection of the fact that for the N=2
superconformal algebra the spectral flows for $\th=\pm 1$ commute with
the twistings.
 In order to be rigorous we must trade the spectral
flow operator $\cU_1$ by an operator, say $\cA$, such that the topological
algebra automorphism can be properly expressed as

\BE\new\BA{rclcrcl}
\cA \, \cL_m \, \cA^{-1}&=& \cL_m - m\cH_m\,,\\
\cA \, \cH_m \, \cA^{-1}&=&-\cH_m - {\ctop\over3} \delta_{m,0}\,,\\
\cA \, \cQ_m \, \cA^{-1}&=&\cG_m\,,\\
\cA \, \cG_m \, \cA^{-1}&=&\cQ_m\,.\
\label{autom} \EA\EE

It is straightforward to prove that $\cA^{-1} = \cA$ and that
$\cA$, like $\cU_1$, maps null vectors into null vectors of the
same level and U(1) charges related by $\htop \leftrightarrow -\htop-\ctop/3$.
Again, this has no direct analog neither in the
N=2 superconformal case nor for the mirror automorphism of the
twisted topological algebra. In fact, we have found the operators
that transform null states built on the chiral ring into both
null states built on the chiral ring and null states built on the
antichiral ring; they will be published in a forthcoming paper
\cite{JIB}.

\section{Level 2 Results}\lvm

Now we will apply the results obtained in the previous section to
the case of level 2 descendants and null vectors. We will adopt
the {\it algebra automorphism} point of view, and use the
transformation \req{autom}. Equivalently, we could have used the
spectral flow mapping \req{spfla} between the two twisted theories,
keeping track of the labels (1) and (2).

Let us start by noticing that any topological descendant $\ket\Ups$
 can be decomposed as the sum of a $\cQ_0$-invariant descendant
 $\ket\Ups^Q$ and a $\cG_0$-invariant descendant $\ket\Ups^G$,
 as deduced from the anticommutator
$\{\cQ_0, \cG_0\} = 2 \cL_0$. For this reason we will concentrate
mainly on those states.

Let us call $\cO^{(q)}$ the operator acting on the primary state
$\ket\P_\htop$, with U(1) charge $\htop$, to build the
descendant $\ket\Ups^{(q)}$, namely $\ket\Ups^{(q)} = \cO^{(q)}
\ket\P_\htop$. The {\it relative} U(1) charge $q$ of
$\ket\Ups^{(q)}$ is the charge carried by $\cO^{(q)}$, given
by the number of $\cG$ modes minus the number of $\cQ$ modes
in each term. The U(1) charge of $\ket\Ups^{(q)}$ is therefore
$\htop + q$.
 We will denote by $\cO^{(q)Q}$ and $\cO^{(q)G}$
the operators corresponding to $\cQ_0$-invariant and
$\cG_0$-invariant descendants, $\ket\Ups^{(q)Q}$ and
$\ket\Ups^{(q)G}$, respectively.
The transformation \req{autom} changes the U(1) charges of
the descendants and primary states in the form
$\htop \leftrightarrow -\htop-{\ctop\over3}$. As a consequence the
 relative charges are mapped as $q \leftrightarrow -q$ and the operators
 of type $\cO^{(-q)Q}$ are transformed into operators of type
$\cO^{(q)G}$.

At level 2 there are five kinds of operators $\cO^{(q)}$
regarding the possible values of $q$ and the invariance
properties under $\cQ_0$ and $\cG_0$ of the corresponding
descendants $\ket\Ups^{(q)}$. These operators are  $\cO^{(-1)Q},
 \cO^{(1)G}, \cO^{(0)}, \cO^{(0)Q}$ and $\cO^{(0)G}$. Notice that only
 for $q=0$ there exist descendants that are neither
$\cQ_0$ nor $\cG_0$ invariant (although a sum of both
kinds, as we mentioned before). The general rule, at any level,
is that the descendants with largest absolute value of $q$
are either $\cQ_0$-invariant (for $q<0)$ or
$\cG_0$-invariant (for $q>0)$.

\vskip .2in

The generic $\cO^{(-1)Q}$ operator can be written as

\BE \cO^{(-1)Q} = \cQ_{-2} + \a \cL_{-1} \cQ_{-1} +
  \b \cH_{-1} \cQ_{-1}  \label{ferQdes} \EE

\noi
The transformed operator reads

\BE \cA \cO^{(-1)Q} \cA= \cG_{-2} + \a \cL_{-1} \cG_{-1} +
  (\a - \b) \cH_{-1} \cG_{-1}   \label{ferGdes} \EE

\noi
and is a generic $\cO^{(1)G}$ operator.
 Null vectors of type $\ket\Ups^{(-1)Q}$ and
 $\ket\Ups^{(1)G} = \cA \ket\Ups^{(-1)Q}$
are given by the two sets of solutions
\BE
\a=\ccases{6\over\ctop-3}{\ctop-3\over6}\!,
\qquad\beta=\ccases{12\over\ctop-3}{\ctop+3\over6}\!,
\qquad\a-\beta=\ccases{6\over3-\ctop}{-1}\!,\EE

\noi
corresponding to the following U(1) charges of the primary states
on which they are built
\BE
\htop^Q=\ccases{1-\ctop\over2}{-{\ctop+3\over2}}\!,
\qquad\htop^G=\ccases{\ctop-3\over6}{1}\!.\EE

\noi
Notice that $\htop^G = - \htop^Q - \ctop/3$.

\vskip .2in

Now let us consider the operators with relative charge $q = 0$.
The general expression for $\cO^{(0)}$ is

\BE \cO^{(0)} = \a \cL_{-1}^2 + \th \cL_{-2} +
 \G \cH_{-1} \cL_{-1} + \b \cH_{-1}^2 +
 \g \cH_{-2} + \delta \cQ_{-1} \cG_{-1}   \label{bosdes} \EE

\noi
$\cQ_0$-invariance of the associated descendant gives the constraints

\BE  \b = 0, \quad \g = 0, \quad \G = 2\delta  \label{Q0inv} \EE

\noi
while $\cG_0$-invariance results in the equations

\BE  2 \a - \G + 2 \delta = 0,\ \ \a -\G + \b = 0, \ \
2 \th + \a -\g + 2 \delta = 0  \label{G0inv} \EE
\noi
The transformed operator is also of generic $\cO^{(0)}$ type and reads

\BE\new\BA{rcl}
 \cA \cO^{(0)} \cA = \a \cL_{-1}^2 &+& (2 \delta + \th) \cL_{-2} +
 (2 \a - \G) \cH_{-1} \cL_{-1} + (\a - \G + \b) \cH_{-1}^2 \\
 &+& (2\th+\a-\g+2\delta) \cH_{-2}
 - \delta \cQ_{-1} \cG_{-1} . \EA \label{bostrf} \EE

\noi
It is straightforward to verify that this operator corresponds to
a $\cG_0$-invariant ($\cQ_0$-invariant) descendant if the former
operator corresponds to a $\cQ_0$-invariant ($\cG_0$-invariant)
descendant.

For $\cQ_0$-invariant null vectors $\ket\Ups^{(0)Q}$, using \req{Q0inv}
and setting $\th=1$, one finds \cite{BeSe2}

\BE
\cO^{(0)Q}=\a\cL_{-1}^2+\cL_{-2}+\Gamma\cH_{-1}\cL_{-1}+\half
\Gamma\cQ_{-1}\cG_{-1}\,,\EE

\noi
with
\BE
\a=\ccases{6\over\ctop-3}{\ctop-3\over6}\!,
\qquad\Gamma=\ccases{6\over3-\ctop}{-1}\!,
\qquad\htop^Q=\ccases{\ctop-3\over6}{1}\!. \label{solfer}\EE

\noi
The transformed null vectors $\ket\Ups^{(0)G}$ given by
the operator \req{bostrf}, are $\cG_0$-invariant with $\htop^G=-\htop^Q-
\ctop/3$. The resulting $\cO^{(0)G}$ is given by
(using \req{Q0inv} and \req{solfer})

\BE
2\delta+\th=\ccases{9-\ctop\over3-\ctop}{0}\!,
\qquad 2\a-\Gamma=\ccases{18\over\ctop-3}{\ctop\over3}\!.\EE
\BE
\a-\Gamma=\ccases{12\over\ctop-3}{\ctop+3\over6}\!,
\qquad 2\th + \a +2\delta =\ccases{2}{\ctop+3\over6}\!,
\qquad\htop^G=\ccases{1-\ctop\over2}{-{\ctop+3\over3}}\!.\EE

\vskip .2in

Observe that $\cO^{(0)Q}$ is much simpler (four terms) than
 $\cO^{(0)G}$ (six terms). The reason is that terms containing
$\cH$ modes alone are absent in $\cQ_0$-invariant descendants
of type $\ket\Ups^{(0)}$ \cite{BeSe3}. This is true at any level, so that
at any level the direct computation of $\cG_0$-invariant
null states $\ket\Ups^{(0)G}$ is much harder than that
of $\cQ_0$-invariant null states $\ket\Ups^{(0)Q}$ (already
at level 2 the difference in computing time for them is one
order of magnitude: roughly two hours for $\ket\Ups^{(0)Q}$ and
about 20 hours for  $\ket\Ups^{(0)G}$ ).
 By using the transformation \req{autom},
however, we obtain straightforwardly null states of type
$\ket\Ups^{(0)G}$ from null states of type $\ket\Ups^{(0)Q}$
(in the example above the 20 hour work reduces to 15 minute work).
\newpage
\section{DDK and KM Realizations of the Topological Algebra}\lvm

In this section we will use the spectral flow mapping \req{spfla}
to relate the DDK and KM realizations of the topological algebra
\req{topalgebra} \cite{BeSe2} \cite{BeSe3} \cite{BJI2}.
 These are the two twistings of the
same N=2 superconformal theory .
The DDK topological field theory, a bosonic string construction
\cite{DDK},
corresponds to the twist (2), while the KM topological field theory,
that is related to the KP hierarchy through the Kontsevich-Miwa
transformation, corresponds to the twist (1).
The field content in these realizations consists of
$d \leq 1$ matter, a scalar and a $bc$ system.
 The scalars differ by the background charge and the
 way they dress the matter $(Q_s = Q_{Liouville},\  \Delta = 1$ in
the DDK case versus $Q_s = Q_{matter},\  \Delta = 0 $ in the KM
case), while the $bc$ systems differ by the spin and the central
charge $(s=2,\  c=-26$ for the DDK ghosts
 versus $s=1,\  c=-2$ for the KM ghosts).

\vskip .2in

In the DDK realization the generators of the topological algebra are

\BE \cL_m^{(2)}=L_m^{(2)}+l_m^{(2)},\qquad
l_m^{(2)}\equiv\sum_{n\in{\oZ}}(m+n):\!b_{m-n}^{(2)}c_n^{(2)}\!:\label{L26}\EE

\BE\cH_m^{(2)}=\sum_{n\in\oZ}:\!b_{m-n}^{(2)}c_n^{(2)}\!:{}-
{}\sqrt{{3-\ctop\over3}}I_m^{(2)}, \qquad\cG_m^{(2)}=b_m^{(2)}
\,,\label{H26}\EE

\BE\cQ_m^{(2)}=2\sum_{p\in\oZ}c_{m-p}^{(2)}L_p^{(2)}
+\sum_{p,r\in\oZ}(p-r):\!b_{m-p-r}^{(2)}c_p^{(2)}c_r^{(2)}\!:{}+
{}2\sqrt{{3-\ctop\over3}}m
\sum_{p\in\oZ}c_{m-p}^{(2)}I_p^{(2)}+
 {\ctop\over3}(m^2-m)c_m^{(2)}~,\label{Q26}\EE

\noi
and the chiral primary states can be written as
$\ket\P^{(2)} = \ket\xi^{(2)} \otimes c_1^{(2)} \ket0_{gh}$,
where $\ket\xi^{(2)}$ is a primary state in the "matter + scalar" sector
(for a spin-2 $bc$ system $c_1 \ket0_{gh}$ is the
 "true" ghost vacuum
annihilated by all the positive modes $b_n$ and $c_n$).

In the KM realization the generators read

\BE\cL_m^{(1)}=L_m^{(1)}+l_m^{(1)},\quad
l_m^{(1)}=\sum_{n\in{\oZ}}n\!:\!b_{m-n}^{(1)}c_n^{(1)}\!: \label{L}\EE

\BE\cH_m^{(1)}= - \sum_{n\in\oZ}:\!b_{m-n}^{(1)}c_n^{(1)}\!: +
{}\sqrt{{3-\ctop\over3}}I_m^{(1)},\qquad \cQ_m^{(1)}=b_m^{(1)}
\,,\label{H}\EE

\BE\cG_m^{(1)}=2\sum_{p\in\oZ}c_{m-p}^{(1)}
L_p^{(1)}+2{}\sqrt{{3-\ctop\over3}}
\sum_{p\in\oZ}(m-p)c_{m-p}^{(1)}I_p^{(1)}
{}+\sum_{p,r\in\oZ}(r-p):\!b_{m-p-r}^{(1)}c_r^{(1)}c_{p}^{(1)}\!:
{}+{}{\ctop\over3}(m^2+m)c_m^{(1)}\,,\label{G}\EE

\noi
and the chiral primary states split as
 $\ket\P^{(1)} = \ket\xi^{(1)} \otimes
\ket0_{gh}$.

\vskip .2in
Let us analyze the spectral flow mapping \req{spfla} at the level
of these specific realizations; in other words, how the spectral
flow maps the different DDK and KM fields into each other.
Starting with the simplest transformation,
$ \, \cU_1 \cG_m^{(2)}\cU_1^{-1} = \cQ_m^{(1)} \, $, we get

\BE \cU_1 \,  b_m^{(2)} \, \cU_1^{-1} = b_m^{(1)}. \label{Ub} \EE

\noi
{ } From the anticommutation relation $\{b_m,c_n\} = \Kr{m+n}$, for
any $bc$ system, and using \req{Ub}, one obtains

\BE \cU_1 \,  c_m^{(2)} \, \cU_1^{-1} = c_m^{(1)}. \label{Uc} \EE

\noi
Now taking into account the ghost vacuum behaviour

\BE b^{(2)}_{n \geq -1} \ket0_{gh} = c^{(2)}_{n \geq 2} \ket0_{gh} =0,
\qquad b^{(1)}_{n \geq 0} \ket0_{gh} =c^{(1)}_{n \geq 1} \ket0_{gh} =0,
\EE

\noi
it is easy to deduce

\BE \cU_1 \sum_{n \in\oZ} :\!b_{m-n}^{(2)}c_n^{(2)}\!: \cU_1^{-1} =
\sum_{n \in\oZ} :\!b_{m-n}^{(1)}c_n^{(1)}\!: - \Kr{m} \EE

\noi
and

\BE \cU_1 \ket0_{gh} = b^{(1)}_{-1} \ket0_{gh}, \qquad
    \cU_{-1} \ket0_{gh} = c^{(2)}_1 \ket0_{gh} .\EE

\noi
{ } From the transformations
$ \, \cU_1 \cH^{(2)}_m\cU_1^{-1}=-\cH_m^{(1)} -{\ctop\over3}
 \delta_{m,0} \, $ and
$ \, \cU_1\cL^{(2)}_m\cU_1^{-1}= \cL_m^{(1)} - m\cH_m^{(1)}$,
one finally obtains

\BE \cU_1 \,  I^{(2)}_m \, \cU_1^{-1}=I_m^{(1)}- \sqrt{3-\ctop\over3}
 \delta_{m,0} \label{UI} \EE

\noi
and

\BE \cU_1 \,  L^{(2)}_m \, \cU_1^{-1}=L_m^{(1)}- m \sqrt{3-\ctop\over3}
 I_m^{(1)} + \delta_{m,0}. \label{UL} \EE

\noi
The second of these expressions, although useful, does not contain
any new information, since $L_m$ contains the matter contribution
(invariant under $\cU_1$), plus the scalar contribution, whose
spectral flow transformation is already given in the previous line.
Thus we see, from \req{Ub}, \req{Uc} and \req{UI}, that the spectral
flow transformation, at the level of the DDK and KM realizations,
does not mix fields of different nature, as it does at the
topological algebra level (mapping $\cQ_m$ and $\cG_m$ into each other).
The mirror map \cite{RS} does in fact mix the matter, the scalar
and the ghost fields. This shows, again, a deep difference between
the spectral flow map and the mirror map between the two twisted
 topological theories corresponding to a given N=2 superconformal
theory.

To finish, let us consider the {\it reduced} DDK and KM conformal
field theories (CFT's). These are the theories given by
the "matter + scalar" systems, without
the ghosts, and are described by the commutation
relations

 \BE\new\BA{rcl}
\L[L_m,L_n\R]&=&(m-n)L_{m+n}+{D\over 12}(m^3 -m)\Kr{m+n}~,\\
\L[L_m,I_n\R]&=&-nI_{m+n}-\half Q_s (m^2+m)\Kr{m+n}~,\\
\L[I_m,I_n\R]&=&-m\Kr{m+n}~.\EA\label{hat}\EE

\noi
where

  \BE  L_m = L_m^{matter}-\half\sum_n:I_{m-n}I_n: +\ \half Q_s(m+1)I_m\EE

\noi
and  $D=26 \  (D=2) ,\   Q_s=\sqrt{{25-d\over3}}
 \  (Q_s=\sqrt{{1-d\over3}})$
for the DDK { }(KM)  realization.

\vskip .2in

Using the spectral flow transformations \req{UI} and \req{UL},
between the DDK and KM CFT's,
 one readily deduces that primary fields, as well as
null vectors, are mapped into each other. This was to be expected
since the spectral flow maps topological $\cG_0$-invariant null
 vectors into $\cQ_0$-invariant null vectors and, at the level of
the DDK and KM realizations, those reduce to ghost-free
null vectors of the DDK and KM CFT's respectively \cite{BJI}.

\section{Final Remarks}\lvm

The main issue in this letter has been to write down and analyze the
spectral flow mapping between the two twisted topological
 theories associated to a given N=2 superconformal theory.
We have found that this mapping has better properties, at the level
of descendant states, than the corresponding spectral flow on
the untwisted N=2 superconformal theories: it preserves the level
of the states and it maps null states into null states (spectral flows
on the N=2 superconformal algebra do not map null states built on
the chiral ring into null states built on the antichiral ring, and
modify the level of the states).
 Exactly
for the same reasons this mapping has also a better behaviour than the
mirror map between the two twisted theories.

We have also found  that the spectral flow mapping interpolating
between the two twisted theories gives rise to a topological
algebra automorphism which acts inside a given theory. This is
a reflection of the fact that for the N=2 superconformal theories
 the spectral flows commute with the twistings. Again, this
automorphism preserves the level of the states, transforming null
states into null states (the mirror map automorphism of the topological
algebra fails to do that).

Both the spectral flow mapping and the automorphism, provide a powerful
tool to compute topological null states, known to be related with
Lian-Zuckermann states and other extra states relevant in string
theory (see, for example \cite{MV}). As an example,
 we have written down all the level 2 results: the different types
of general topological descendants with their spectral flow
transformations and the specific coefficients which correspond to
null states. We show that difficult null states can be computed
straightforwardly from much easier ones.

We think, however, that the most interesting use of the spectral
flow mapping written down here
 should be found at the level of the specific realizations of the
 topological algebra. We have just started this program,
analyzing the behavior of the DDK and KM realizations of the
topological algebra under the spectral flow action. We have found that,
contrary to what happens with the mirror map, the spectral flow
does not mix the different component fields (matter, scalar and ghosts).
As a result, the null states of the reduced conformal field theories
(without ghosts) are mapped into each other too.

 In general, it may well happen that the image theory
of a physically relevant theory is much simpler
to deal with than the latter (like seems to be the case
for the DDK and KM theories). In this respect let us remember
that {\it almost all string theories, including the bosonic
string, the superstring, and W-string theories, possess a
twisted N=2 superconformal symmetry} \cite{BLNW}.
There are several other interesting twisted N=2 topological
theories, like the one considered in \cite{MV} for $d=1$ string
theory, or those related to the Kodaira-Spencer theory of
gravity \cite{BCOV}, or those considered in \cite{RS},
 etc (just to mention a few). It would be interesting to investigate
whether the spectral flow mapping described here has something
useful to add to the known results about all or some of those
 topological theories.

\vskip 1cm

\centerline{\bf Acknowledgements}

We would like to thank Bert Schellekens
 for valuable discussions and for carefully reading the manuscript.


\begin{thebibliography}{9}
\def\NPB{Nucl. Phys. B}
\def\PLB{Phys. Lett. B}
\def\MPLA{Mod. Phys. Lett. A}

\bibitem{[EY]} T.~Eguchi and S.~K.~Yang, \MPLA5 (1990) 1653

\bibitem{[W-top]} E.~Witten, Commun. Math. Phys. 118 (1988) 411;
 \NPB340 (1990) 281

\bibitem{DVV} R. Dijkgraaf, E. Verlinde and H. Verlinde, \NPB352
(1991) 59

\bibitem{SS} A. Schwimmer and N. Seiberg, \PLB184 (1987) 191

\bibitem{LVW} W.~Lerche, C.~Vafa and N.~P.~Warner,
 \NPB324 (1989) 427

\bibitem{RS} A.V. Ramallo and J.M. Sanchez de Santos, "Topological
Matter, Mirror Symmetry and Non-critical (Super)Strings",
hep-th/9505149 (1995)

\bibitem{JIB} B. Gato-Rivera and J.I. Rosado, work in preparation

\bibitem{BeSe2} B.~Gato-Rivera and A.~M.~Semikhatov, \PLB293 (1992) 72,
 Theor. Mat. Fiz. 95 (1993) 239, Theor. Math. Phys. 95 (1993) 536

\bibitem{BeSe3} B.~Gato-Rivera  and A.~M.~Semikhatov, \NPB408 (1993) 133

\bibitem{BJI2} B.~Gato-Rivera and J.~I.~Rosado, "Topological Theories
from Virasoro Constraints on the KP Hierarchy". Talk given at the
"28th International Symposium on the Theory of Elementary Particles",
Wendisch - Rietz (Germany), August 1994, hep-th/9411185

\bibitem{DDK} F.~David, \MPLA3 (1988) 1651;\\
 J.~Distler and H.~Kawai, \NPB321 (1989) 509

\bibitem{BJI} B.~Gato-Rivera and J.~I.~Rosado, \PLB346 (1995) 63

\bibitem{BLNW} M. Bershadsky, W. Lerche, D. Nemeschansky and
N.P. Warner, \NPB401 (1993) 304

\bibitem{MV} S. Mukhi and C. Vafa, \NPB407 (1993) 667

\bibitem{BCOV} M. Bershadsky, S. Cecotti, H. Ooguri and C. Vafa,
Commun. Math. Phys. 165 (1994) 311

\end{thebibliography}
\end{document}